\documentclass[epj]{webofc}
\usepackage[varg]{txfonts}   
\usepackage{ctable}
\usepackage{siunitx}

\newcommand{\pt} {\mbox{$p_{\rm T}$}}
\newcommand{\RAA}{R_{\rm AA}}

\woctitle{Quarks-2016}
\begin{document}
\title{Upgrade of the ALICE Inner Tracking System}
%
%

\author{
  \firstname{Iouri} \lastname{Belikov}\inst{1}
  \fnsep\thanks{\email{iouri.belikov@iphc.cnrs.fr}}
  \firstname{}
  \lastname{for the ALICE Collaboration}
}

\institute{IPHC, Universit\'{e} de Strasbourg, CNRS-IN2P3, 23 rue du Loess,
  BP28, 67037 Strasbourg cedex 2}

\abstract{%
A Large Ion Collider Experiment (ALICE) is built to study
the properties of the strongly interacting matter created in heavy-ion
collisions at the LHC. With the upgrade of its Inner Tracking System (ITS),
the ALICE experiment is going to increase the rate of data taking by almost
two orders of magnitude. At the same time, the precision of secondary vertex
reconstruction will become by at least a factor 3 better than it currently is.
In this talk, we briefly show some selected physics results motivating
the upgrade of the ITS, describe the design goals and the layout
of the new detector, and highlight a few important measurements
that will be realized after the completion of this upgrade.  
}
\maketitle
%

\section{ALICE experiment at CERN}
\label{intro}
A Large Ion Collider Experiment (ALICE)~\cite{Carminati:2004fp,Alessandro:2006yt,ALICEjinst} is one of the four major experiments
currently operating on the beams of the Large Hadron Collider (LHC) at the
European Organization for Nuclear Research (CERN).
The aim of this experiment is to study the physics of strongly
interacting matter at extreme energy densities, where the formation of
a new phase of matter, the quark-gluon plasma, is expected.
For this purpose, the collaboration is carrying out series of comprehensive
measurements of hadrons, electrons, muons and photons produced in
the collisions of heavy nuclei. In addition, ALICE is studying
proton-proton interactions
both as a reference for the heavy-ion collisions and also in physics areas
where ALICE is competitive with other LHC experiments.

\section{Selected (Run I) results and their limitations}
An important part of the ALICE physics program is dedicated to high precision
measurements of charm and beauty production in heavy-ion collisions.
The ultimate goals of the heavy-flavour studies include:
\begin{itemize}
\item Thermalization of heavy quarks in the produced medium (by measuring
  baryon-to-meson ratios for charm and beauty particles, ratios of yields
  of heavy-flavour particles with strange content, and azimuthal flow
  anisotropy for as many as possible heavy-flavour species).
\item Parton mass and colour-charge dependence of in-medium energy losses (by
  measuring the momentum-dependent nuclear modification factors for B and D
  mesons, and comparing them with those for light-flavour particles).
\end{itemize}

\begin{figure}[htb]
\centering
\includegraphics[width=0.43\textwidth]{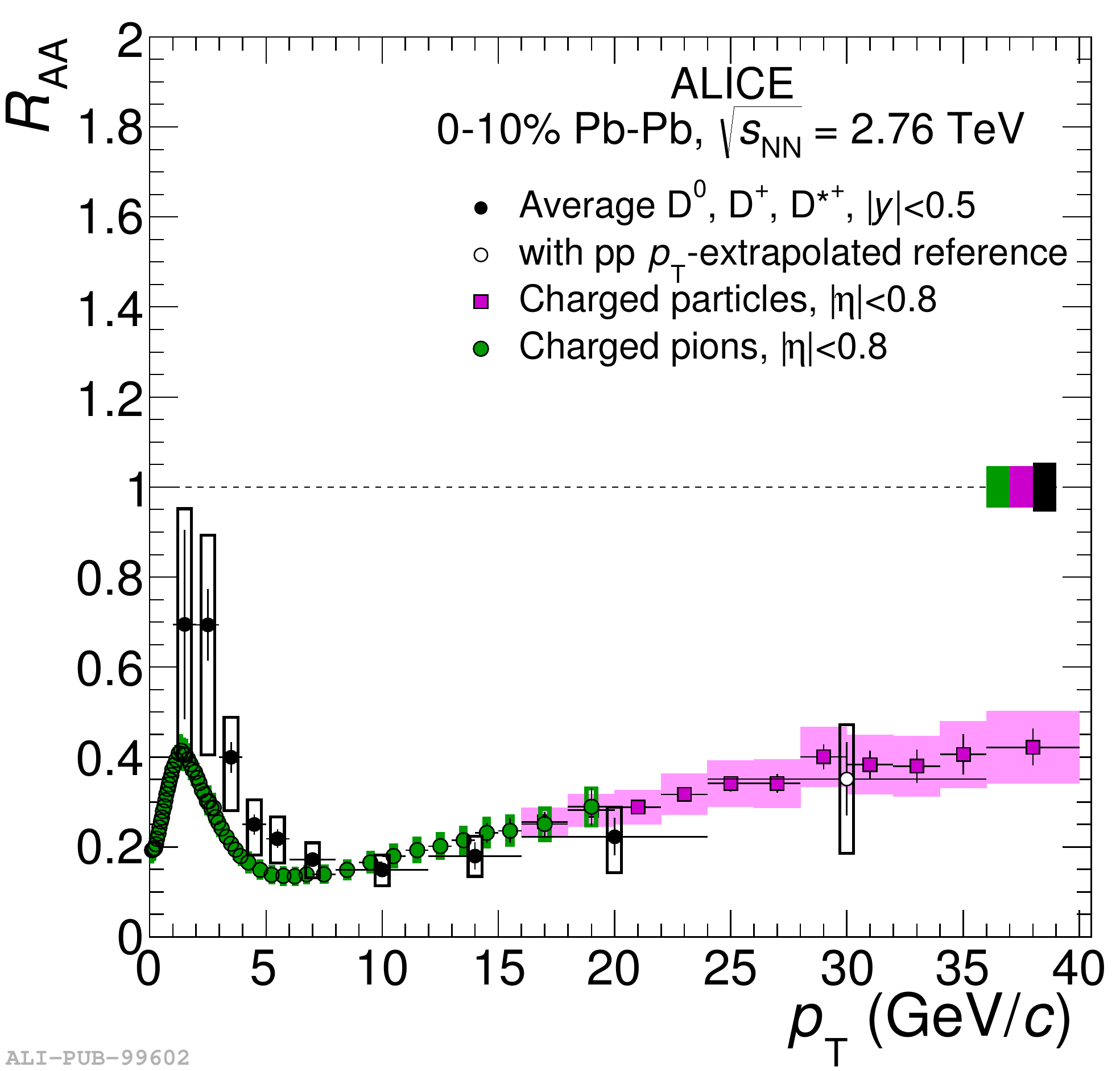}
\includegraphics[width=0.55\textwidth]{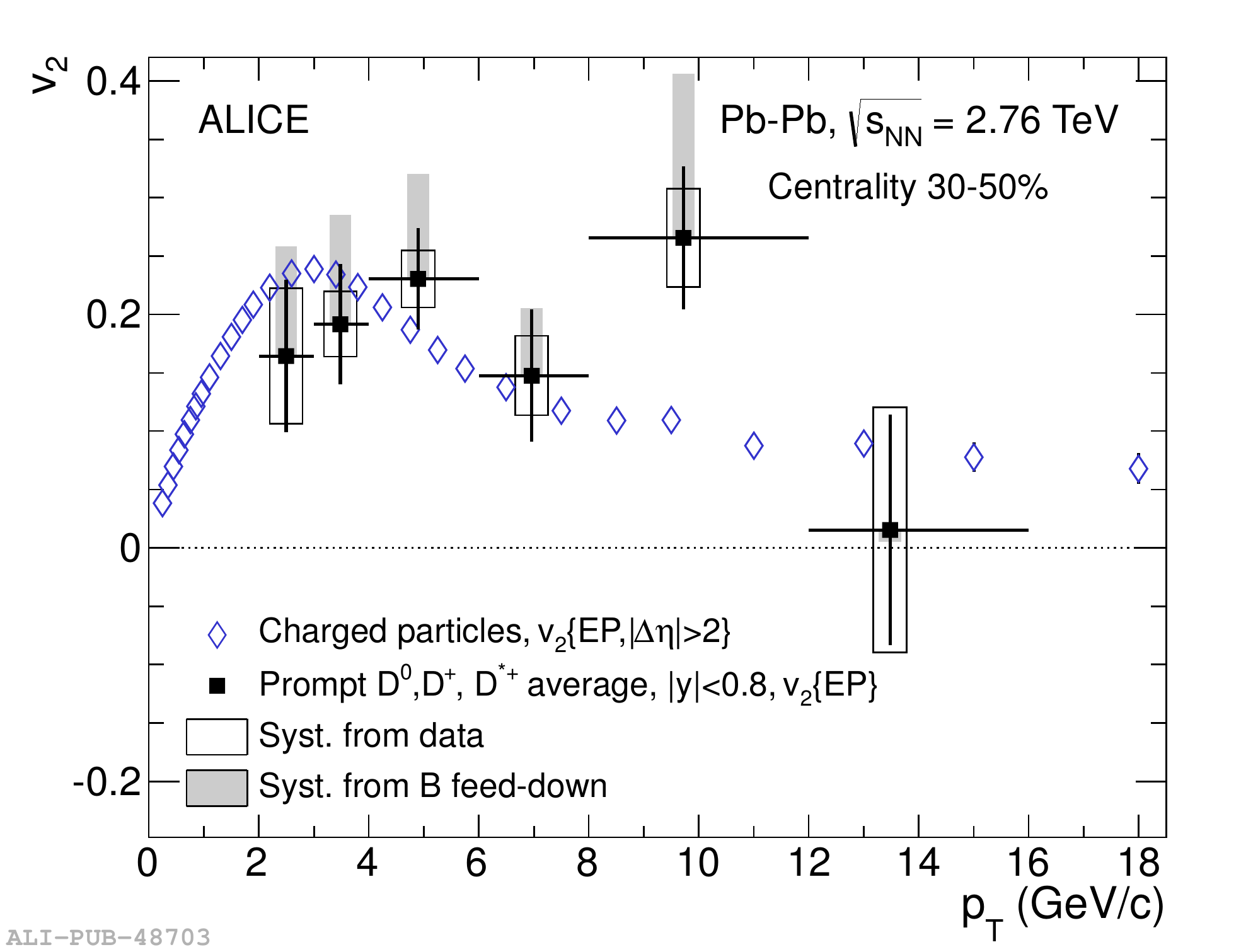}
\caption{Left panel: Nuclear modification factor
  $\RAA$ for D-mesons, pions and charged particles~\cite{Adam:2015sza}.  Right panel: Amplitude of the
  second-order flow harmonic $v_2$ for D-mesons and charged particles~\cite{Abelev:2013lca}.
  Statistical and systematic uncertainties are represented by bars
  and boxes respectively.
}
\label{raa_v2}       
\end{figure}

Using the data recorded during the LHC Run I (2009--2013),
the ALICE Collaboration has already published several important results
on the D-meson production in Pb--Pb collisions. 
For example, the left panel of Figure~\ref{raa_v2} shows the
$\pt$-dependent nuclear modification factor for D-mesons
(average of D$^0$,  D$^+$ and D$^{*+}$) compared to that
of pions and charged particles in the 0--10~\% centrality class.
There is an indication that production of D-mesons in Pb--Pb collisions at
lower $\pt$ is less suppressed than that of light-flavour particles.
The ALICE measurement of the second-order azimuthal flow harmonic $v_2$
is presented on the right panel of the Figure. The data suggest that the
$v_2$ for D-mesons is compatible with that of charged particles and is
larger than 0 over a wide momentum range. But, in both cases,
the current statistical and systematic uncertainties do not allow for firm
conclusions. In addition, the read-out rate capabilities and space-point
precision of the present Inner Tracking System (ITS) are not sufficient
to perform similar measurements with beauty particles (B-mesons), which are
important to prove the existence of quark-mass ordering:
$\RAA^{\rm B} > \RAA^{\rm D} > \RAA^{\rm charged}$ and
$v_2^{\rm B} < v_2^{\rm D} < v_2^{\rm charged}$.

At the same time, even though ALICE has already measured several
baryon-to-meson ratios with light-flavour
particles~\cite{Abelev:2013xaa,Adam:2015kca}, the extension
of these results towards the heavy-flavous sector is currently not possible.
The read-out rates and precision of track reconstruction provided
by the present ITS do not allow for detecting
the heavy-flavous baryons ($\Lambda_{\rm c}$ and $\Lambda_{\rm b}$)
in high-track-multiplicity environment of Pb--Pb collisions at the LHC.

\section{Upgrade of the Inner Tracking System}
\begin{figure}[htb]
  \centering
\includegraphics[width=0.77\textwidth]{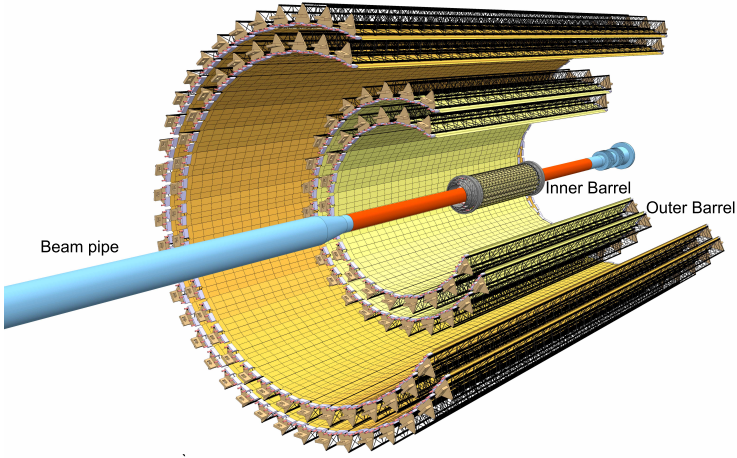}
\caption{The general layout of the upgraded ITS detector.}
\label{layout}       
\end{figure}

The limitations of the present ITS will be radically reduced with
the planned upgrade~\cite{Abelevetal:2014dna}.
The general layout of the upgraded ITS
is shown in Figure~\ref{layout}.
The detector will be 1.5 m long, with the outer radius of 40 cm,
and the overall surface of about 10~m$^2$ will be covered with
silicon pixel sensors. This will be a ``12.6 Giga-pixel camera'',
with an estimated cost of about 13.6 million Swiss Francs. 

The detector will consist of seven layers,
with the first layer as close to the interaction point as 22.4~mm.
The material budget of the three innermost layers will be as low as
0.3~\% of a radiation length. Combined with the space-point precision
of about 5~$\mu$m, the expected track impact-parameter
resolution at $\pt\sim$1~GeV/$c$ will be about 20~$\mu$m, which is a factor 3
(in the transverse plane) and 5 (in the beam direction) better
as compared with the current situation.

The new detector will also allow for an efficient track reconstruction
down to very low $\pt$.
The simulations show that the tracking efficiency at $\pt\sim$0.1~GeV/$c$
will be a factor about 6 higher than it is now. This will be of a great
importance for planned di-electron measurements (see Table~\ref{tableReach})
and the detection of D$^*$ and D$_{\rm S}$ mesons.

All seven layers of the upgraded ITS will be equipped with the ALPIDE
chip~\cite{Abelevetal:2014dna}. Taking advantages of the
0.18~$\mu$m CMOS technology by TowerJazz~\cite{towerjazz}, this chip
combines the sensitive part and the read-out electronics within the same
piece of silicon.
The main characteristics of the chip are:
\begin{itemize}
  \item overall chip size: $15\times 30$~mm$^2$;
  \item pixel size: $29\times 27~\mu$m$^2$;
  \item detection efficiency: $>99$~\%, with a
  noise probability of $<10^{-6}$;
  \item time resolution: about $2~\mu$s;
  \item in-pixel discriminators and in-matrix address encoder with asynchronous sparsified readout;
  \item power consumption: 40~mW/cm$^2$;
  \item high radiation tolerance: tested up to 2.7 Mrad (TID), and
    $1.7\times 10^{13}$~1~MeV~n$_{\rm eq}$/cm$^2$ (NIEL)
\end{itemize}

The upgraded ITS will be almost two orders of magnitude faster than now.
The read-out electronics of the detector will be able to register
events with a typical rate of 50~kHz and a few 100~kHz for minimum
bias Pb--Pb and pp collisions respectively.

The parts of the detector will have to be aligned with the precision
of better than 5~$\mu$m by means of dedicated software. The data coming
out of the detector will be reconstructed quasi-online, using
a dedicated Online-Offline (O$^2$) computer farm~\cite{O2}.

Overall, the upgrade of the Inner Tracking System will be accomplished
by the end of the LHC Long Shutdown 2 (2021).

\section{Physics with the upgraded ITS}
 
The expected physics reach for various
observables is summarized in Table~\ref{tableReach},
in terms of minimum accessible $\pt$ and of statistical uncertainties.
We consider a scenario with an integrated luminosity of
\SI{10}{\per\nano\barn}, fully used for minimum-bias Pb--Pb data collection,
and a low-magnetic-field run with \SI{3}{\per\nano\barn}
of integrated luminosity for the low-mass di-electron studies.
The case of the programme up to Long Shutdown 2 is shown for comparison.
In this case, a delivered luminosity of
\SI{1}{\per\nano\barn} is assumed, out of which 10\% is recorded with a
minimum-bias trigger.

In the upgrade case, the systematic
uncertainties will be reduced as well. This is because for many
measurement, due to the mentioned improvements in the track impact-parameter
resolution, the level of the combinatorial background will significantly be
be reduced, and the feed-down contribution from beauty to charm measurements
will be estimated directly, from the data themselves.

There are also several observables, like those involving the heavy-flavour
baryons and the low-momentum di-electrons, that were never accessible before,
but will become well within the experimental reach thanks to the unique
capabilities of the upgraded ALICE Inner Tracking System.

\begin{table}[htb]
\centering
\caption{Summary of the physics reach: minimum accessible $\pt$ and relative
  statistical uncertainty in Pb--Pb collisions for an integrated luminosity of
  \SI{10}{\per\nano\barn}. For heavy flavour, the statistical uncertainties
  are given at the maximum between $\pt=\SI{2}{GeV/\it{c}}$ and $p_{\rm T}^{\rm min}$.
  For elliptic flow measurements, the value of $v_2$ used to calculate
  the relative statistical uncertainty $\sigma_{v_2}/v_2$ is given
  in parenthesis. The programme up to Long Shutdown 2, with an integrated
  luminosity of \SI{0.1}{\per\nano\barn} collected with minimum-bias trigger,
  is shown for comparison.}
\label{tableReach}
\begin{tabular}{lcccc}
\toprule        
& \multicolumn{2}{c}{Current, \SI{0.1}{\per\nano\barn} } & \multicolumn{2}{c}{Upgrade, \SI{10}{\per\nano\barn}
} \\
            \cmidrule{2-3} \cmidrule{4-5}
Observable & $p_{\rm T}^{\rm min}$  & \multicolumn{1}{c}{statistical} & $p_{\rm T}^{\rm min}$  & \multicolumn{
1}{c}{statistical} \\
            & (GeV/$c$)  & \multicolumn{1}{c}{uncertainty} & (GeV/$c$)  & \multicolumn{1}{c}{uncertainty} \\
\midrule
             \multicolumn{5}{c}{Heavy Flavour} \\
\midrule
D meson $R_{\rm AA}$           & 1   & \SI{10}{\percent}  &  0 & \SI{0.3}{\percent}\\
D$_{\rm s}$ meson $R_{\rm AA}$ & 4 & \SI{15}{\percent} &  $<2$ & \SI{3}{\percent}\\
D meson from B $R_{\rm AA}$ & 3 & \SI{30}{\percent} &  2 & \SI{1}{\percent}\\

J/$\psi$ from B $R_{\rm AA}$ & 1.5 & ~~~\SI{15}{\percent} \tiny{($\pt$-int.)}
&  1 & \SI{5}{\percent}\\

B$^+$ yield & \multicolumn{2}{c}{not accessible} & 2    & \SI{10}{\percent} \\
$\Lambda_{\rm c}$ $\RAA$ & \multicolumn{2}{c}{not accessible} & 2 & \SI{15}{\percent}\\
$\Lambda_{\rm c}/{\rm D^0}$ ratio & \multicolumn{2}{c}{not accessible} &  2 & \SI{15}{\percent} \\
$\Lambda_{\rm b}$ yield & \multicolumn{2}{c}{not accessible} & 7 & \SI{20}{\percent} \\
D meson $v_2$ ($v_2 = 0.2$) &  1 & \SI{10}{\percent}  &  0 & \SI{0.2}{\percent} \\
D$_{\rm s}$ meson $v_2$ ($v_{2}=0.2$) & \multicolumn{2}{c}{not accessible} &  $<2$ & \SI{8}{\percent}\\
D from B $v_2$ ($v_2 = 0.05$) & \multicolumn{2}{c}{not accessible}  &  2 & \SI{8}{\percent}\\
J/$\psi$ from B $v_2$ ($v_2 = 0.05$) & \multicolumn{2}{c}{not accessible} &  1 & \SI{60}{\percent}\\
$\Lambda_{\rm c}$ $v_2$ ($v_2 = 0.15$) & \multicolumn{2}{c}{not accessible}  &  3 & \SI{20}{\percent} \\
\midrule
             \multicolumn{5}{c}{Di-electrons} \\
\midrule
Temperature (intermediate mass)  &  \multicolumn{2}{c}{not accessible}  &       & \SI{10}{\percent}  \\
Elliptic flow ($v_2 = 0.1$)~\cite{LOI}  &  \multicolumn{2}{c}{not accessible}     &       & \SI{10}{\percent}  \\
Low-mass spectral function~\cite{LOI}  &  \multicolumn{2}{c}{not accessible}     &  0.3  & \SI{20}{\percent}  \\
\midrule
             \multicolumn{5}{c}{Hypernuclei} \\
\midrule
$^{\,3}_{\Lambda}$H yield & 2 & 18\,\% &  2  & \SI{1.7}{\percent} \\
\bottomrule
\end{tabular}
\end{table}

%
%
%

\end{document}